# Low-Cost Optoacoustic Tomography System with Programmable Acoustic Delay-Line

Daohuai Jiang, Hengrong Lan, Yiyun Wang, Feng Gao, and Fei Gao, *Member, IEEE*

***Abstract*—Photoacoustic tomography (PAT) is an emerging technology for biomedical imaging that combines the superiorities of high optical contrast and acoustic penetration. In the PAT system, more photoacoustic (PA) signals are preferred to be detected from full field of view to reconstruct PA images with higher fidelity. However, the requirement for more PA signals' detection leads to more time consumption for single-channel scanning based PAT system, or higher cost of data acquisition (DAQ) module for an array-based PAT system. To address this issue, we proposed a programmable acoustic delay line module to reduce DAQ cost and accelerate imaging speed for PAT system. The module is based on bidirectional conversion between acoustic signals and electrical signals, including ultrasound transmission in between to provide sufficient time delay. The acoustic delay line module achieves tens or hundreds of microseconds' delay for each channel, and is controlled by a programmable control unit. In this work, it achieves to merge four inputs of PA signals into one output signal, which can be recovered into original four PA signals in the digital domain after DAQ. The imaging experiments of pencil leads embedded in agar phantom is conducted by the PAT system equipped with the proposed programmable acoustic delay-line module, which demonstrated its feasibility in biomedical imaging system.**

## I. Introduction

Photoacoustic tomography (PAT), as a kind of emerging noninvasive imaging modality, has shown its unique advantages in biomedical imaging, combining the strength of both deep penetration of ultrasound wave and high contrast of optical absorption [1-5]. In a conventional PAT system, after excitation of a laser pulse, an ultrasonic transducer array will receive the PA signals from different directions and reconstruct the PA image from these PA signals [6]. The more PA signals acquired for image reconstruction, the less distortion and artifacts for the reconstructed PA image. To accelerate the PAT imaging speed or get more PA signals at one laser shot, the ultrasound array with more elements is applied for PA signals detection in the PAT system [7-9]. However, more PA signals' detection will consume more high-speed data acquisition (DAQ) channels, leading to dramatic cost increase for PAT system [10].

For a full-ring ultrasound transducer array-based PAT system, the useful PA signal length depends on the diameter of the ring transducer. It is interesting to observe that the PA signals' durations are usually less than 40 microseconds (assuming that the velocity of the PA wave is 1500 m/s and the diameter is 60 mm, then $t=(60mm)/(1500m/s)=40\mu s$) with laser pulse repetition rate of 20 Hz, which means that the time interval between two PA signals is about 50 milliseconds [11]. Therefore, it turns out that for most of the operating time (1-40μs/50ms = 99.92%), the status of DAQ is idle. Fig. 1(a) shows that the conventional PA signal sampling method with PA signal connecting with DAQ one by one, whose DAQ work status is shown in Fig. 1(c). In each pulse cycle, one channel of DAQ only samples one PA signal, and for most of the time in one cycle, DAQ works at idle status. To utilize the DAQ's capability more efficiently and reduce the number of acquisition channels for PA signals' sampling, we proposed the method of time-sharing multiplexing for PA signals sampling by applying a custom-designed programmable acoustic delay-line (PADL) module in this paper. Fig. 1(b) shows the *n*-in-1 data sampling method with *n*-in-1 PADL module, where *n* channels' PA signals are sampled only by single channel DAQ. Fig. 1(d) shows the DAQ working status: at one pulse cycle, one channel of the DAQ can sample *n* channels' PA signals with *n* times higher use efficiency.

The time-sharing multiplexing method depends on the realization of analog signals' time delay. Murat Kaya Yapici, et al proposed a parallel acoustic delay line for PAT based on the detection and ultrasound transmission by optical fiber, which however suffers limited sensitivity and bandwidth [12, 13]. The delay line module based on capacitor storage in [14] is complex for delayed signals' recovery. To realize tens or hundreds of microseconds' time delay for analog signal, the mostly used method is via the acoustic wave propagation. The acoustic wave that travels at velocity $v$ at a distance $d$ gives $t_d=d/v$ time delay

This research was funded by Natural Science Foundation of Shanghai (18ZR1425000), and National Natural Science Foundation of China (61805139). (Corresponding author: Fei Gao.)

Daohuai Jiang (e-mail: jiangdh1@shanghaitech.edu.cn), Hengrong Lan (e-mail: lanhr@shanghaitech.edu.cn) and Yiyun Wang (e-mail: wangyy@shanghaitech.edu.cn) are with the Hybrid Imaging System Laboratory (HISLab), Shanghai Engineering Research Center of Intelligent Vision and Imaging, School of Information Science and Technology, ShanghaiTech University, Shanghai 201210, China, with Chinese Academy of Sciences, Shanghai Institute of Microsystem and Information Technology, Shanghai 200050, China, and also with University of Chinese Academy of Sciences, Beijing 100049, China.

Feng Gao and Fei Gao are with the Hybrid Imaging System Laboratory (HISLab), Shanghai Engineering Research Center of Intelligent Vision and Imaging, School of Information Science and Technology, ShanghaiTech University, Shanghai 201210, China (e-mail: gaofeng@shanghaitech.edu.cn, gaofei@shanghaitech.edu.cn).

[15]. Generally speaking, the requirements for a dedicated acoustic delay line module include low acoustic attenuation over a wide bandwidth, smaller velocity (smaller size for same delay time), easy signal recovery after propagation, proper signal amplitude compensation, compatibility for different transducer types, etc.

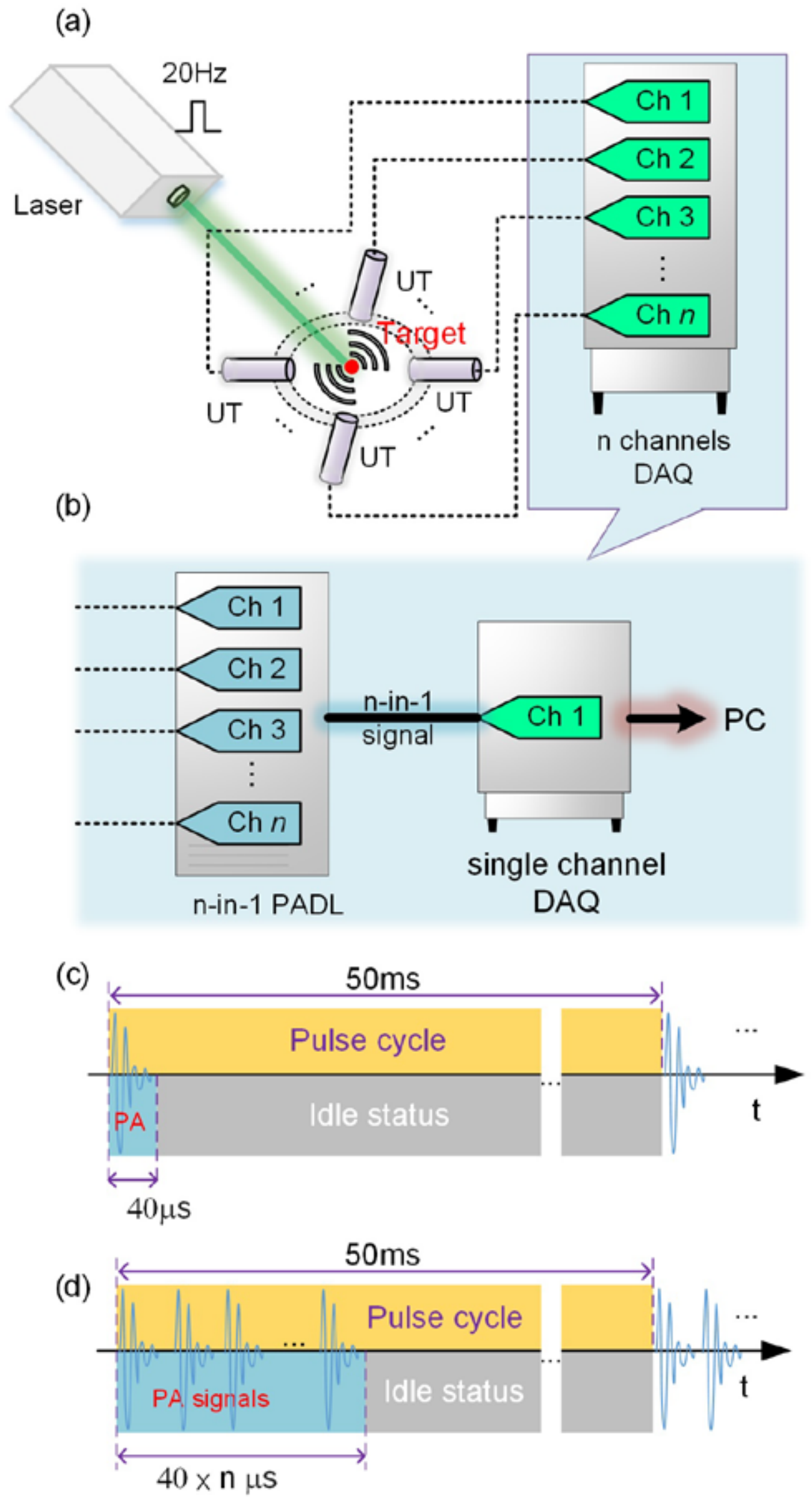

Fig. 1. The PA signals sampling setup with (a) conventional method and (b) time-sharing multiplexing by a PADL module; (c) and (d) are the DAQ working status corresponding to (a) and (b) setup. UT: ultrasound transducer; DAQ: data acquisition; PADL: programmable acoustic delay-line; PC: personal computer.

In this paper, to address the abovementioned issues, we proposed a four-in-one programmable acoustic delay-line (PADL) module for PA signal acquisition. This module can achieve time delay ranging from tens to hundreds of microseconds, including a programmable unit to achieve different time delays. By exactly recovering the original PA signals without time delay, image reconstruction will be performed in the digital domain after DAQ sampling.

This paper is organized as follows: Section II presents the methodology of the delay line module including the general wave extrapolation and delayed signal recover method. In section III, a PAT system based on the PADL module will be introduced, and a pencil leads made phantom is imaged to verify the feasibility of the delay line module. Section IV presents discussions and conclusions for the PADL module based PAT system.

## II. Programmable Acoustic Delay-Line

The method of reducing DAQ cost for PAT system is based on the delay line module to delay the signals in analog domain and recover the signals after sampling in digital domain. In this section, we will introduce the programmable acoustic delay line module, the general wave extrapolation and delayed signal recovery method primarily.

### A. PA Physical Theory

The PA wave equation is shown below describing the PA wave propagation in both time and spatial domains [16, 17]:

$$\left(\nabla^2 - \frac{1}{c^2}\frac{\partial^2}{\partial t^2}\right)p(r,t) = 0 \tag{1}$$

subject to the initial conditions:

$$p(r,t)\big|_{t=0} = \frac{\beta c^2}{C_p}A(r); \qquad \frac{\partial p(r,t)}{\partial t}\Big|_{t=0} = 0, \tag{2}$$

where $\nabla^2$ denotes the Laplacian operator, and $\boldsymbol{A}(\mathbf{r})$ is the distribution of absorbed optical energy density. The constants $\beta$, $c$ and $C_p$ denote the thermal coefficient of volume expansion, speed of sound and the specific heat capacity of the medium at constant pressure, respectively. Assuming that the received photoacoustic wave is $p(r,t)$, the PA image is reconstructed by PA signals which reflect the $p(r,t)$ information [18]. The relationship of delayed PA signals we received and $p(r,t)$ function can be regarded as:

$$s(i,t+t_d) \propto p(r,t+t_d) \tag{3}$$

where $p(r,t+t_d)$ is delay acoustic pressure corresponding to $i^{th}$ delayed PA signal $s(i,t+t_d)$, $t_d$ is delay time. The delayed signal $s(i,t+t_d)$ can be recovered into $s(i,t)$ without time delay. Moreover, the signal $s(i,t)$ is corresponding to the acoustic pressure of $p(r,t)$. Hence, with proper data processing, the delay signals $s(i,t+t_d)$ can also reconstruct the PA image.

### B. Delay-line Design and Implementation

Achieving nanoseconds' time delay for analog signal is available by extended electric wire or by some commercial integrated chip. However, the PA signal's duration is dozens of microseconds, which requires at least tens or hundreds of microseconds' delay time for the time-division multiplexing sampling method. However, the conventional method based on electrical or optical signal time delay is difficult to achieve tens of microseconds' level [14]. On the other hand, the acoustic propagation velocity is several orders of magnitude slower than the speed of electrons. Therefore, we propose to transform the electrical analog signal to acoustic signal, which can easily achieve tens or hundreds of microseconds time delay. Fig. 2 shows the 4-in-1 programmable acoustic delay-line module.

The acoustic delay-line unit is based on bidirectional conversion between acoustic signals and electrical signals. Two ultrasound transducers are used for signal conversion, one for signal transmitting and one for receiving. We selected purified water as the acoustic transmission medium, and the acoustic speed in the water at room temperature (25°C) is about 1.5 mm/μs [17, 19]. It deserves noting that water gives less acoustic attenuation and better coupling with the ultrasound transducer. Fig. 2(a) shows the water-made acoustic delay-line structure and its photograph. The distance between two ultrasound transducers is 60 mm, which corresponds to 40 μs time delay ((60mm)/(1.5mm/μs)=40 μs). A transparent polyvinyl chloride (PVC) tube is used to seal water in it. The external diameter of the PVC tube is 15 millimeters and the internal diameter is 13 millimeters. Besides, it is noticed that the PVC tube is fully filled with purified water without air to make better acoustic coupling and propagation.

The above-mentioned delay-line unit in Fig. 2(a) can only achieve a fixed time delay (40 μs), which is insufficient for multi-channel data acquisition that requires different time delays for each channel. To address this issue, based on the delay-line unit, we designed the PADL unit, which can realize programmable time delay. Fig. 2(b) shows the structure and its signal transmission diagram. The acoustic delay-line unit is the basics of the PADL, and three analog switches are used to implement signal control. Besides, a low noise pre-amplifier (LNPAmp) and a variable gain amplifier (VGAmp) are used to compensate for the delayed signal's attenuation. An analog adder connects the delayed signal and delay-line unit input that constitutes close-loop feedback, within which PA signal can run repeatedly to achieve enough time delay (multiple of 40 μs). The three analog switches controlling signal input, signal output and signal feedback are corresponding to $SW_{in}$, $SW_{out}$ and $SW_{fb}$ in the figure, respectively. Besides, the analog switches controlling logics are generated by a field-programmable gate array (FPGA) with a particularly designed time sequence. To make sure the feedback loop works stable, the loop gain including LNPAmp and VGAmp should be less than one.

The schematic diagram of the 4-in-1 delay-line module is shown in Fig. 2(c). The module includes three delay units (1,2, and 3) based on the proposed PADL module, and a transfer unit (T unit) that consists of resistors network. The signal input to T unit experiences no time delay, but the resistor network adjusts the amplitude, and an analog switch ($SWT_{out}$) controls the signal output. Moreover, the other three input signals go through delay units with different time delays. In addition, a four-in-one analog adder combines the four signals into one output. Hence, the four parallel pulse inputs are converted into one pulse train with efficient time intervals, so that the four inputs have no aliasing, then it is able to be recovered into four signals after DAQ sampling in the digital domain.

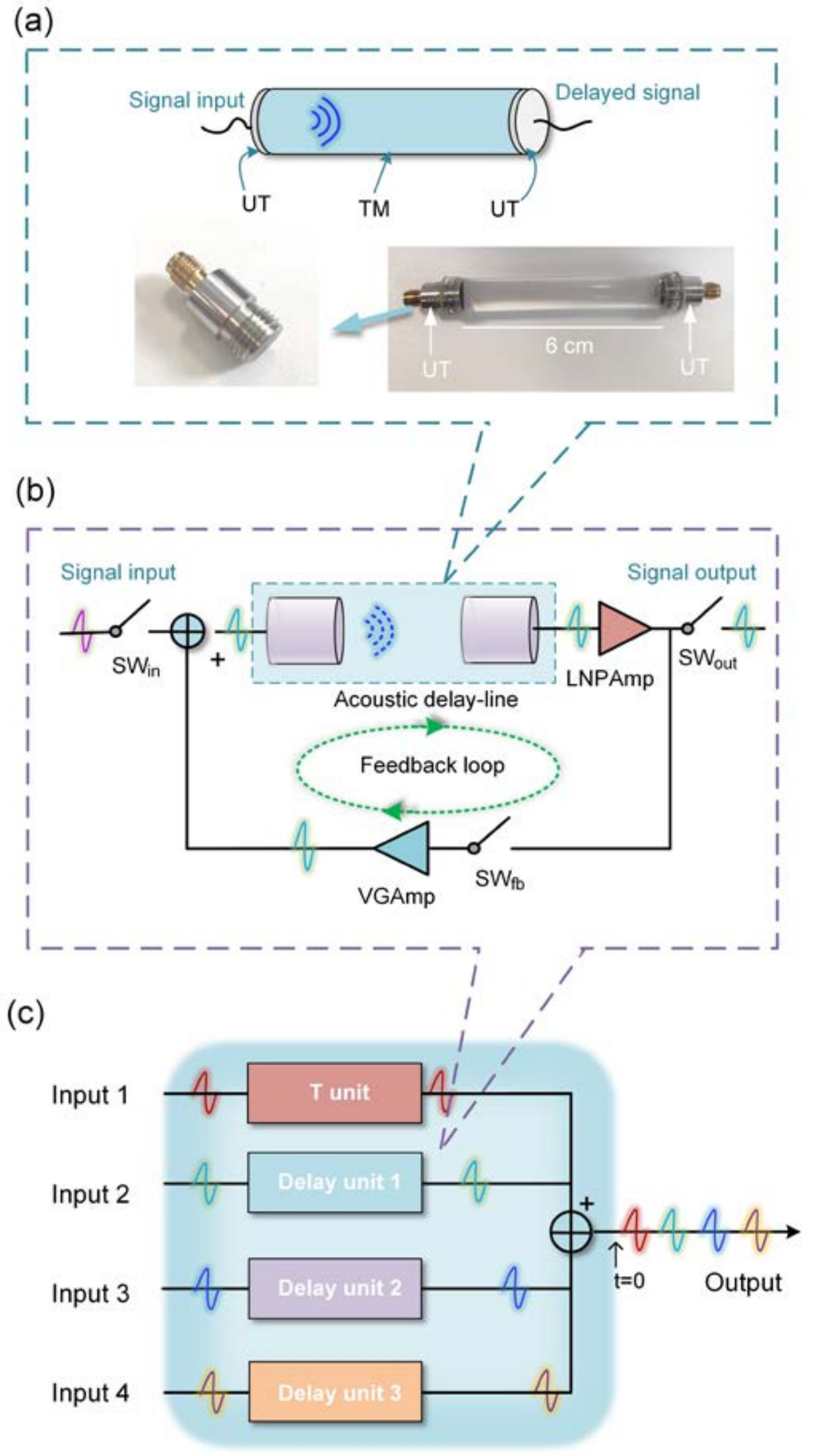

Fig. 2. The structure of the 4-in-1 programmable acoustic delay-line module. (a) the acoustic delay-line unit structure, and its photograph; (b) the schematic of the programmable acoustic delay-line unit; (c) the schematic diagram of the 4-in-1 PADL delay-line module. UT: ultrasound transducer; TM: transmission medium; SW: switch; LNPAmp: low noise pre-amplifier; VGAmp: variable gain amplifier; T unit: transfer unit.

The acoustic delay-line signal transfer characteristics is tested and shown in Fig. 3. Fig. 3(a) shows the acoustic signal waveforms before and after an acoustic delay-line unit. The red curve is the input PA signal detected by the ultrasound transducer. The delayed signal (blue curve) gone through the delay-line achieves 40 μs time delay and amplitude attenuation. The waveform of the output is highly correlated with input signal's waveform. Fig. 3(b) shows acoustic delay-line transfer characteristics with multi-cycle loops. The red curve is the input of the PADL module, and the blue curve of a pulse train is delayed signal output at different cycles. Each pulse of the output signal has interval of 40 μs, so that we can selectively achieve the total delay time by setting proper cycle number.

Except time delay, the PADL module also induces amplitude attenuation that has to be compensated. Fig. 3(c) shows the delayed signal amplitude attenuation, which is related with the distance between two UT and feedback loop cycles. In this design, the selected PVC tube is 60 mm long that causes acoustic attenuation of about 41 dB. The blue star-labeled line stands for the feedback loop cycles caused attenuation, which almost can be neglected.

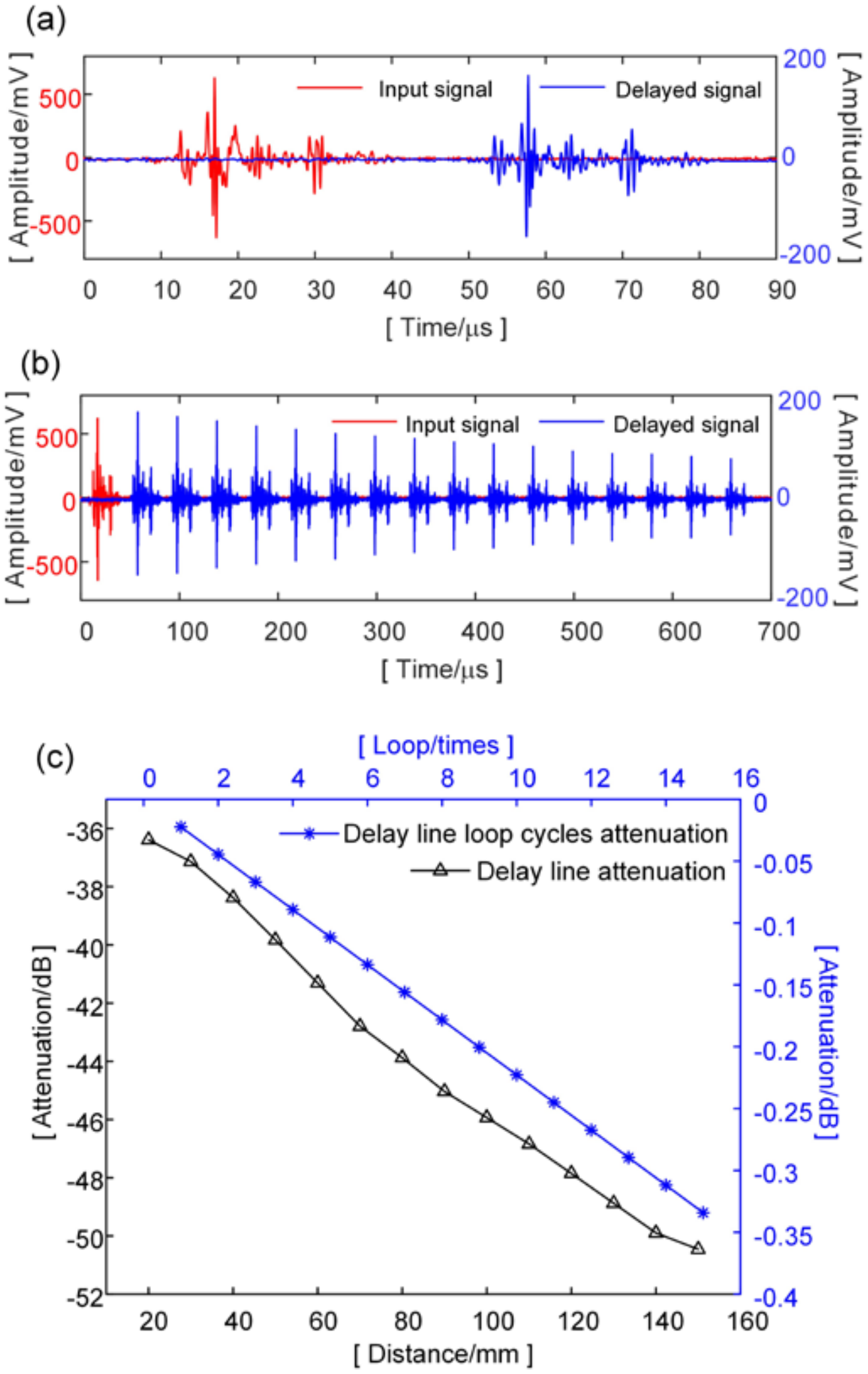

Fig. 3. The signal transfer characteristics of the programmable acoustic delay-line module. (a) acoustic delay-line transfer characteristics of PA signals; (b) acoustic delay-line transfer characteristics with multi-cycle loops; (c) acoustic delay-line signal attenuation characteristics related with UT distance and cycle number.

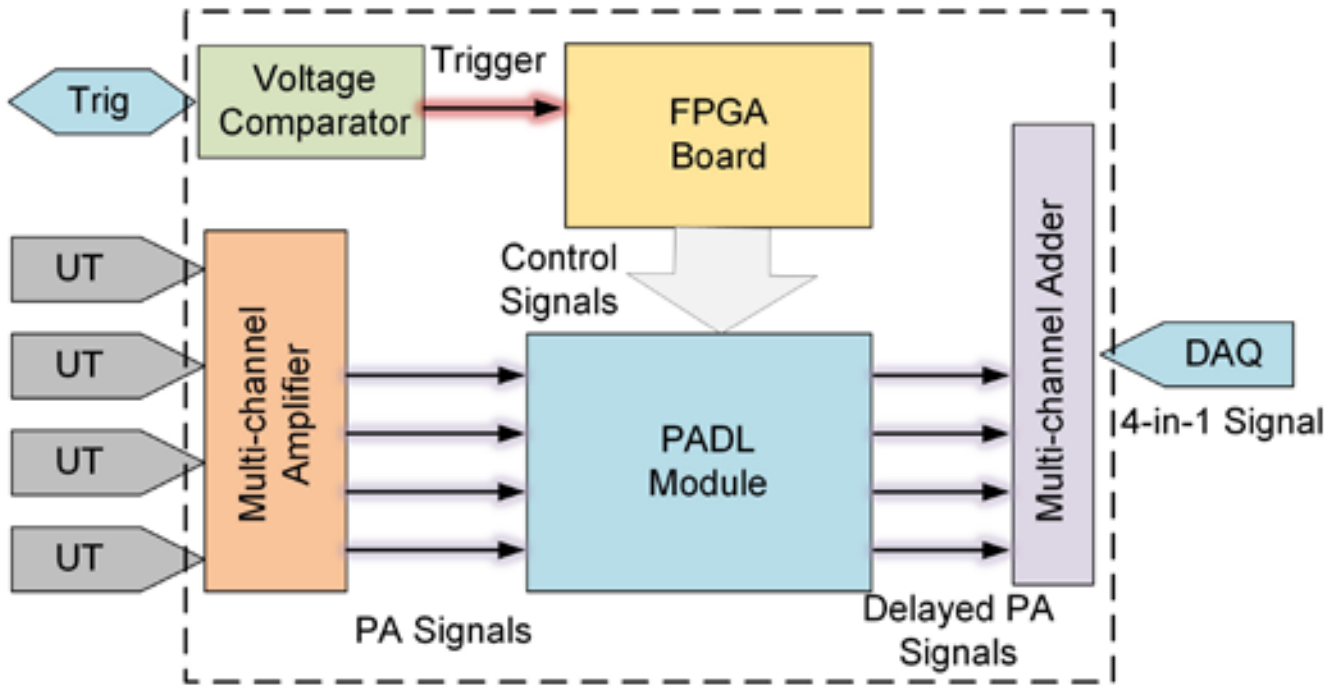

Fig. 4. The 4-in-1 PADL module implementation. Trig: laser trigger; FPGA: field-programmable gate array; UT: ultrasound transducer; PA: photoacoustic; PADL: programmable acoustic delay-line; DAQ: data acquisition device.

### *C. Four-in-one PADL Module Design and Implementation*

The 4-in-1 PADL module contains five parts: multi-channel amplifier, voltage comparator, FPGA board, PADL and multi-channel adder as shown in Fig. 4. The voltage comparator is used for laser trigger detection that generates a trigger signal shown in Fig. 5(a), to FPGA for controlling the PADL. Besides, the FPGA board generates specific control logics to the analog switches of the PADL as shown in Fig. 5(c). The PADL delays the UT detected pre-amplified PA signals, and all the delayed signals enter into the multi-channel adder to generate the 4-in-1 delayed signal that is captured by DAQ.

As shown in previous Fig. 2(b), the analog switches $SW_{in}$, $SW_{out}$ and $SW_{fb}$ control the delay time of the PADL. To achieve different time delay by the 4-in-1 delay-line module, we need to control 10 analog switches in total: a switch for T unit and 3 switches for each delay-line unit. Fig. 5 shows the 4-in-1 PADL module control logic and its corresponding delayed PA signals. Fig. 5(a) is the laser trigger signal: the laser is triggered at the rising edge $t_0$. Fig. 5(b) shows the four PA signals detected by UTs. The detected signals are disturbed by a strong electromagnetic coupling interference signal at $t_0$, which is useless for PA image reconstructions. Hence, the delayed PA signal should exclude the coupling interference signals, so the $SW_{in}$ needs to be closed at $t_1$ that is after $t_0$ and before PA signal comes in. Fig. 5(c) shows the analog switches control logics. Delay units signal inputs switch $SW_{in}$ and T unit output switch $SWT_{out}$ are closed from $t_1$ to $t_2$ that covers the duration containing PA signals, and output the no-delay PA signal (PA1). The delay unit 1 output switch $SW_{out}1$ is closed from $t_2$ to $t_3$, so that it outputs one-cycle delayed signal (PA2). For the delay unit 2, the feedback switch $SW_{fb}$ *Unit2* is closed from $t_2$ to $t_3$ and the output switch $SW_{out}2$ is closed from $t_3$ to $t_4$ , so that it outputs two-cycles delayed signal (PA3). The delay unit 3 feedback switch $SW_{fb}$ *Unit3* is closed from $t_2$ to $t_4$ and the output switch $SW_{out}3$ is closed from $t_4$ to $t_5$, so that it outputs 3-cycles delayed signal (PA4). Fig. 5(d) shows delayed signals' outputs, the blue (PA1), red (PA2), yellow (PA3) and purple (PA4) from the up to down correspond to the output of *T Unit*, *Delay Unit 1*, *Delay Unit 2* and *Delay Unit 3* in Fig. 2(c), respectively. After we get PA signals with different time delays that is long enough to separate these signals in time domain, the multi-channel analog adder combines the four delayed signals into one output. Fig. 5(e) shows the 4-in-1 delayed signals, where the parallel 4 inputs are converted into a serial output. Therefore, by applying the 4-in-1 PADL module, a one-channel DAQ device is available to capture 4 channels' PA signals at one laser shot.

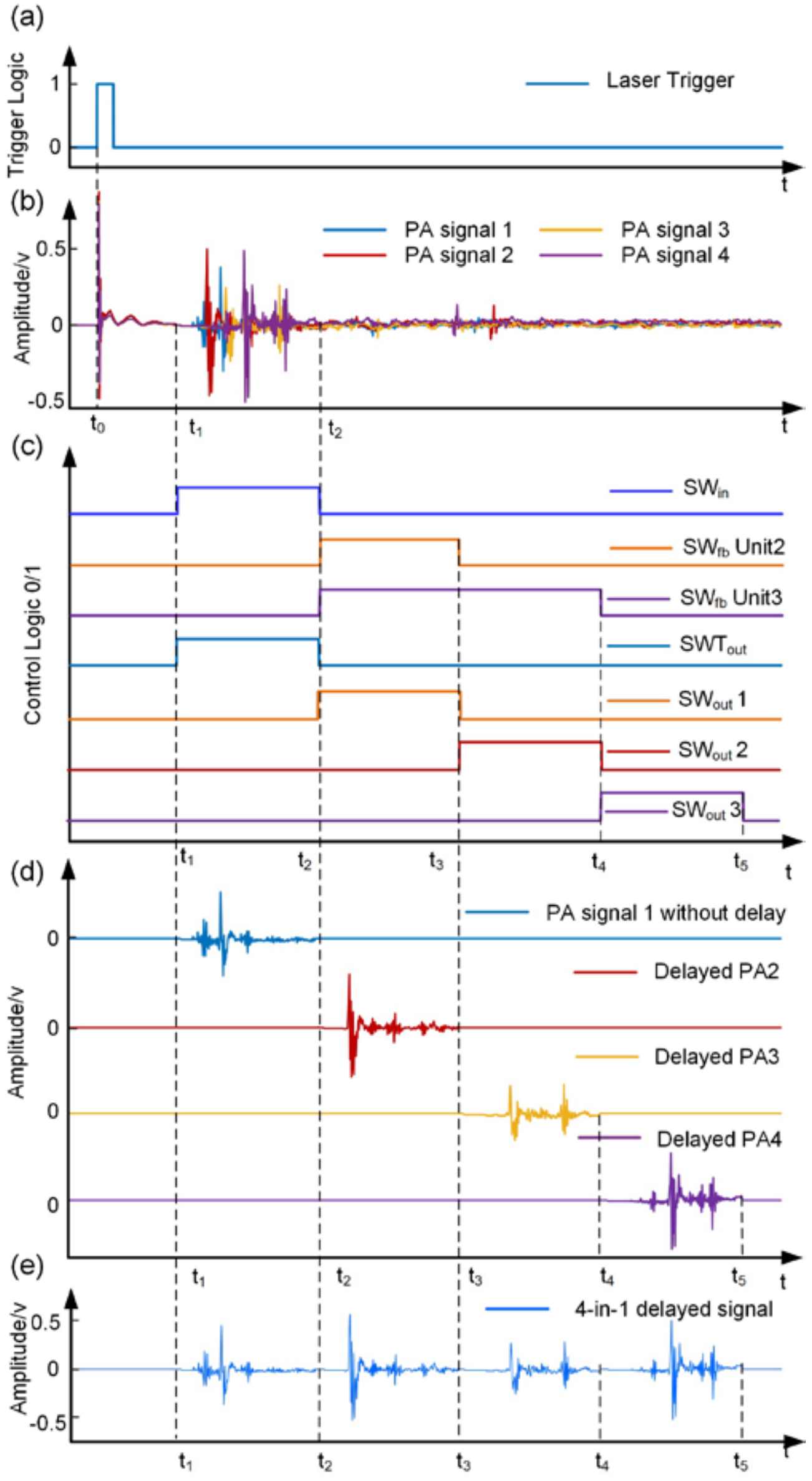

Fig. 5. The 4-in-1 PADL module's control logic waveforms and its corresponding delayed signals and 4-in-1 signal. (a) laser trigger signal; (b) PA signals; (c) the PDLA control logic; (d) delayed signal outputs; (e) the delayed 4-in-1 signal;

## *D. Delayed Signal Recovery*

After we get the 4-in-1 delayed signal, it is feasible to recover the delayed signal into four separate PA signals in digital domain. The signal recovery process in digital domain is the reverse operation of the signal delay in analog domain [20]. Fig. 6 shows the diagram of the 4-in-1 signal' delay and recovery processing. The operation of signal summation, time delay and amplification in analog domain are corresponding to the signal separation, time shift and magnification in digital domain.

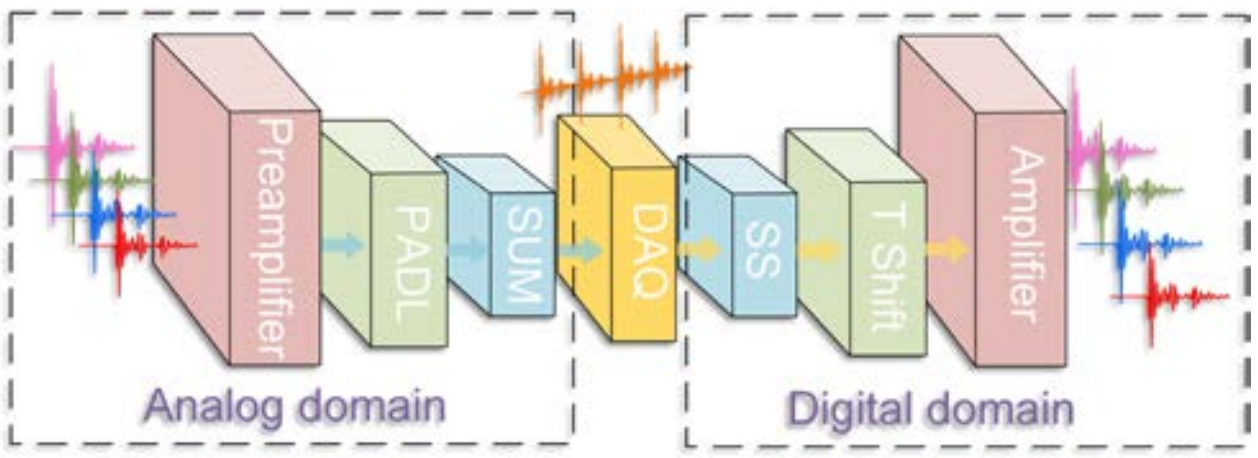

Fig. 6. The diagram of four-in-one signals' delay and recovery processing. PADL: programmable acoustic delay-line module; SUM: summation; DAQ: data acquisition; SS: signal separation; T shift: time shift transform.

The PADL module transferring 4 PA signals with different time delays also induces different amplitude attenuations shown in Fig. 3(c). The signal magnification in digital domain should compensate for this amplitude attenuation. For the signal delay part, the parallel inputs are firstly transformed into a serial signal and sampled by DAQ. After sampling, the 4-in-1 signal should be separated into four parts according to their different delay times, respectively. The parallel input signals and the delayed signal follow below equation:

$$s\left(t+t_{di}\right)=A_i\cdot\sum_{i=1}^{4}\bar{s}\left(i,t\right)\cdot k\left(i,\bar{s}\right) \tag{4}$$

where $s(t+t_{di})$ is delayed signal, $t_{di}$ stands for channel $i$ delayed time of the 4-in-1 PADL module. $\bar{s}(i,t)$ is the parallel input signals, where $i$ ranges from 1 to 4. $k(i,s)$ is a transfer function of a bandpass filter, which is due to the limited bandwidth of UT's bidirectional conversion between acoustic signals and electrical signals of the delay-line module. $A_i$ is the acoustic signals attenuation coefficient and it is determined by:

$$A_i=A_p\cdot A_l\cdot A_m \tag{5}$$

where $A_m$ is the loop gain of each delay-line units. $A_p$ and $A_l$ are acoustic signal attenuation coefficients corresponding to acoustic propagation distance and feedback loops, respectively. $A_p$ can be calculated by the formula below:

$$p_x=p_0e^{-\alpha fx}=p_0\cdot A_p \tag{6}$$

where $p_x$ is the ultrasound amplitude received at a distance $x$, $p_0$ is the ultrasound amplitude at the transmit UT, $\alpha$ is acoustic attenuation coefficient, $f$ is the frequency of the ultrasound signal. To simplify the expression, we define $A_p=exp(-\alpha fx)$ for each channel of the delay line module.

Based on Eq. (4), we can get the related expression in the digital domain:

$$S\left[n+q_i\right]=A_i\cdot\sum_{i=1}^{4}\bar{S}\left[i,n_i\right]\cdot K\left[i,\bar{S}\right]$$
$$n=\left\{1,2,3,\cdots,S_n-q_4\right\},n_i=\left\{1,2,3,\cdots,N_i\right\} \tag{7}$$

where $S[n+q_i]$ is data of the delayed signal and the data length is $S_n$, $n$ ranges from 1 to $S_n$-$q_4$ ($q_4$ *is data length corresponding* *to the longest delay time of the delay-line we set).* $\overline{S}[i,n_i]$ is the data corresponding to input $i$ signal, $n_i$ ranges from 1 to $N_i$ *($N_i$ is the data length of input i)*. $A_i$ and $K[i, \overline{S}]$ are the attenuation coefficient and transfer function of the digital domain. $q_i$ is the data length of channel $i$ corresponding to signal delayed time $t_{di}$. The relationship of $q_i$ and $t_{di}$ with sampling rate $f_s$ can be expressed as:

$$q_i = t_{di} \cdot f_s \tag{8}$$

The signals before delay can be recovered from the sampled signal $S[n+q_i]$, the recovered input $i^{th}$ signal $\overline{S}[i,n_i]$ can be derived as:

$$\begin{gathered} \overline{S}\left[i,n\right] = A_i^{-1} \cdot S\left[n+q_i\right] \cdot K^{-1}\left[i,\overline{S}\right] \\ n = \left\{1,2,3\ldots N_i\right\}, i = \left\{1,2,3,4\right\} \end{gathered} \tag{9}$$

where $A_i^{-1}$ is the amplitude compensation coefficient which is the reciprocal of $A_i$. $K^{-1}[i, \overline{S}]$ is a band-stop filter to compensate UT transfer signal caused by bandpass effect. Hence, by applying above approach, the delayed signal can be recovered without time delay. Furthermore, by applying this module to PAT system, the cost of DAQ device and the whole imaging system can be dramatically reduced.

## III. Imaging Experiment

To verify the feasibility of the 4-in-1 PADL module in PAT system, we demonstrate a four-UT PAT system with a 4-in-1 PADL module. In addition, traditional DAQ sampling method without using PADL module will also be performed for comparison.

### *A. PAT System with PALD Module Setup*

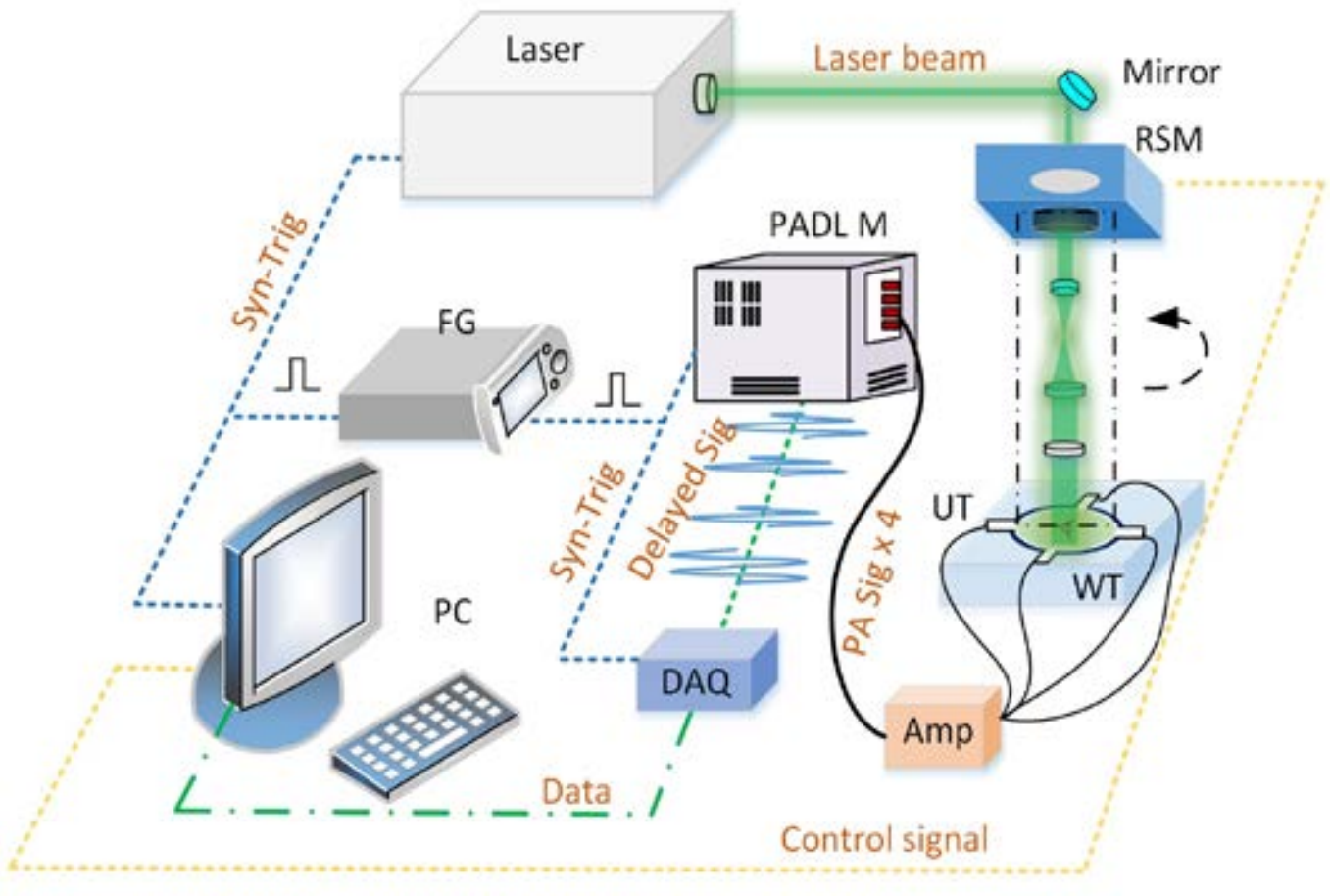

Fig. 7. The PAT system setup with a 4-in-1 PADL module. UT: ultrasound transducer; WT: water tank; Amp: amplifier; PADL M: programmable acoustic delay-line module; Sig: signal; PC: personal computer; DAQ: data acquisition; FG: function generator; RSM: rotational step motor; Syn-Trig: synchronizing trigger.

Fig. 7 shows the PAT system setup with a 4-in-1 PADL module. The laser source (CNI Inc., China) is a Nd:YAG laser with 20 mJ pulse energy, whose wavelength is 532 nm with 10 Hz repetition rate. The optical path is adjusted by a series of optical components like mirror, concave lens and ground glass that guarantees the uniform light irradiation on the imaging phantom. Four single-element unfocused UTs (I2.25P5NF-H, Doppler Inc. China) with 2.25 MHz center frequency are used for PA signal detection. The four detectors are arranged in a concentric circle of 10.6 cm diameter with 90 degrees' separation. A multi-channel low-noise pre-amplifier (PhotoSound Inc. FLASH-AMP16t, USA) with 40 dB gain is used for PA signal amplification. The 4-in-1 PADL module connected the four PA signals after the preamplifier and input the 4-in-1 delayed signal to a DAQ device. It is worth noting that this setup needs only one channel DAQ to sample the 4-in-1 delayed signal. In this experiment, the DAQ device is an oscilloscope (MSO5800, Tektronix Inc.) and working at 1.25 GHz sampling rate. A function generator is used to synchronize the laser, DAQ, PC and 4-in-1 PADL module, and the trigger signal is 10 Hz that is limited by the laser. The personal computer (PC) receives the data from DAQ, and controls the rotational step motor (RSM). The RSM rotation leads the four UTs to move around the imaging phantom. One laser shot corresponds to one-angle PA signal sampling for each UT. Hence, the system moving 90 degrees can achieve a full angle of 360 degrees' PA signals sampling. In this experiment, the step angle is 1 degree.

To get a better signal transformation and recovery, the parameters setup of 4-in-1 PADL module is matching to the PAT system setup. The UTs (I2.25P5NF-H, Doppler Inc. China) used in acoustic delay-line with 2.25 MHz center frequency is used for both acoustic signal detection and acoustic transmit. A variable gain amplifier (AD8368, Analog Devices Inc.) with 0-32 dB gain is used for the delay-line feedback amplitude compensation that adjusts the loop gain. The analog switch of the delay-line unit is a single-pole single-throw analog switch (TS3A4741, Texas Instruments Inc.) with low ON-state resistance ($R_{on}$) and fast switching speeds *($t_{on}$=14 ns, $t_{off}$=9 ns)* that can stably and efficiently control the signals.

When the computer received the data of a 4-in-1 delayed signal from DAQ, the signal recovery processing is performed concurrently. The high-power nanosecond pulsed laser for PAT usually works at a low repetition rate, such as 10 Hz. Therefore, there are about 100 milliseconds to receive and recover the PA signals, which is sufficient for the computer to process the data in the digital domain. Consequently, the PADL module based PAT system with one-channel DAQ can capture four PA signals concurrently, compared with one-channel DAQ without PADL that only captures one PA signal at one laser pulse.

### *B. Delayed Signal Processing*

Based on the above experiment setup, we get the 4-in-1 delayed PA signals of a pencil leads phantom. Fig. 8(a) shows all the 4-in-1 delayed signals. From this figure, we can easily distinguish 4 parts of PA signals in each 4-in-1 delayed signal with different time intervals. The PA signals received by the UT in different angels show different amplitudes. Generally, PA signals with more time delay show more attenuation in amplitude. Moreover, at the time of the laser bursting out *(t=0)*, there are strong coupling signals detected by the UTs that do not alias with the delayed signals due to the delayed signals input to the PADL module cutting off the coupling signals. Next, by applying the proposed method, we can get the recovered PA signals after removing time delay. Fig. 8(b)-(e) show recovered PA signals from one of the 4-in-1 delayed signals, which will be used for image reconstruction.

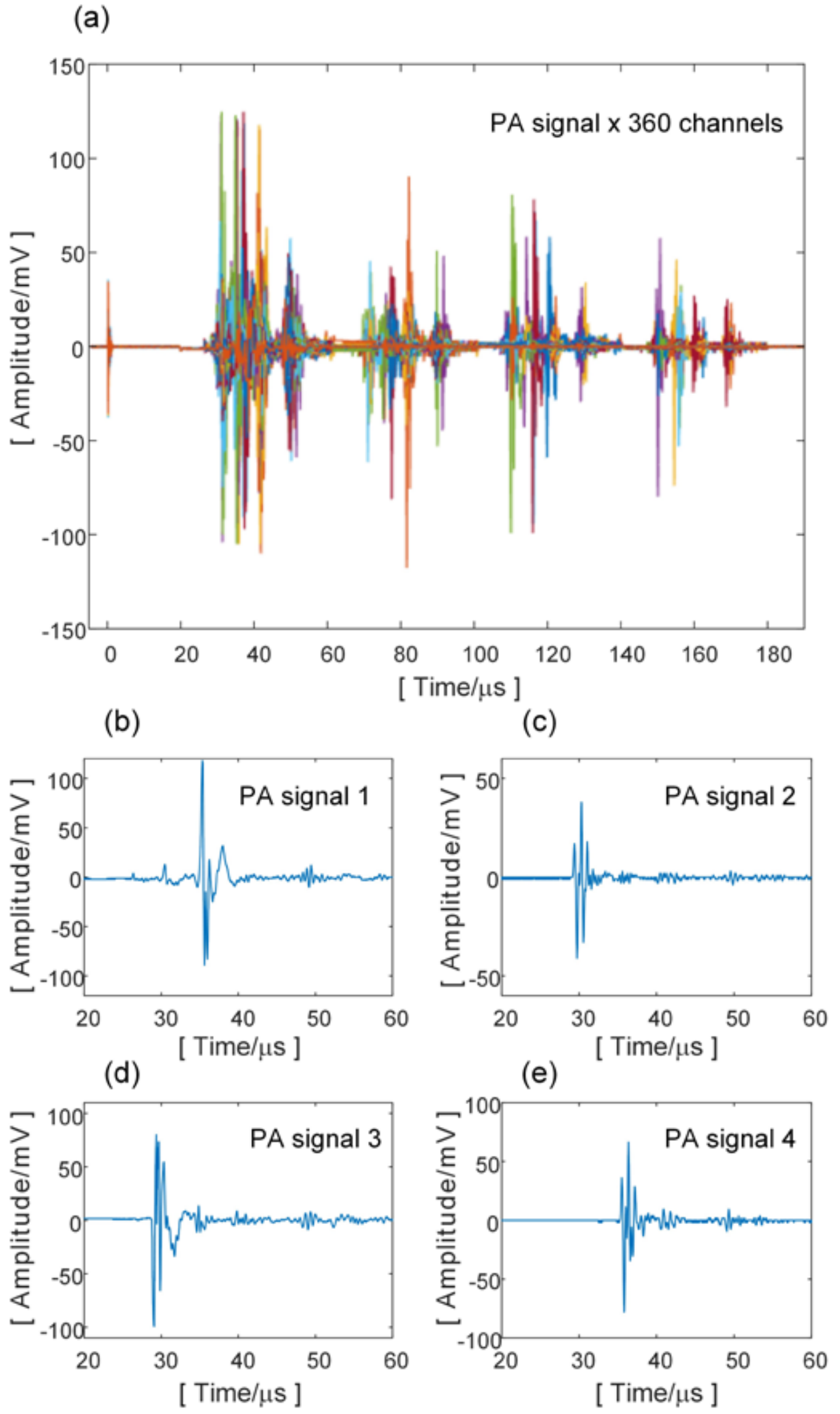

Fig. 8. The 4-in-1 delayed PA signals and recovered PA signals. (a) 360 channels of 4-in-1 delayed PA signals; (b), (c), (d) and (e) is a group of recovered PA signals corresponding to one of 4-in-1 delayed signals.

## C. *Imaging Result*

To demonstrate the feasibility of the 4-in-1 PADL module integrated in the PAT system, we have imaged a pencil leads made phantom with both the traditional sampling method and the proposed sampling method using the 4-in-1 PADL module. The phantom is made up of three crossed 0.5 mm diameter pencil leads embedded in an agar block with inclination. Fig. 9(a) shows the photograph of phantom in top view, whose profile is like a triangle. Moreover, Fig. 9(b) is the photograph of the phantom in lateral view showing the imaging target in different depth, where the white dotted circles indicate the corresponding part between Fig. 9(a) and (b). Fig. 9(c) shows the PAT imaging result by conventional method, and Fig. 9(d) is the imaging result using 4-in-1 delayed signals based PAT image reconstruction. The image reconstruction is implemented in *MATLAB 2018b* version with the classical back-projection algorithm [21, 22]. Both Fig. 9(c) and (d) show the detailed structure of the phantom. The dotted line circled in the phantom is the pencil leads embedded deeper in the agar block, which shows weaker PA signal due to more acoustic attenuation. Overall, it shows that the delayed signals suffers negligible distortion and achieve even better imaging results with lower system cost than conventional method.

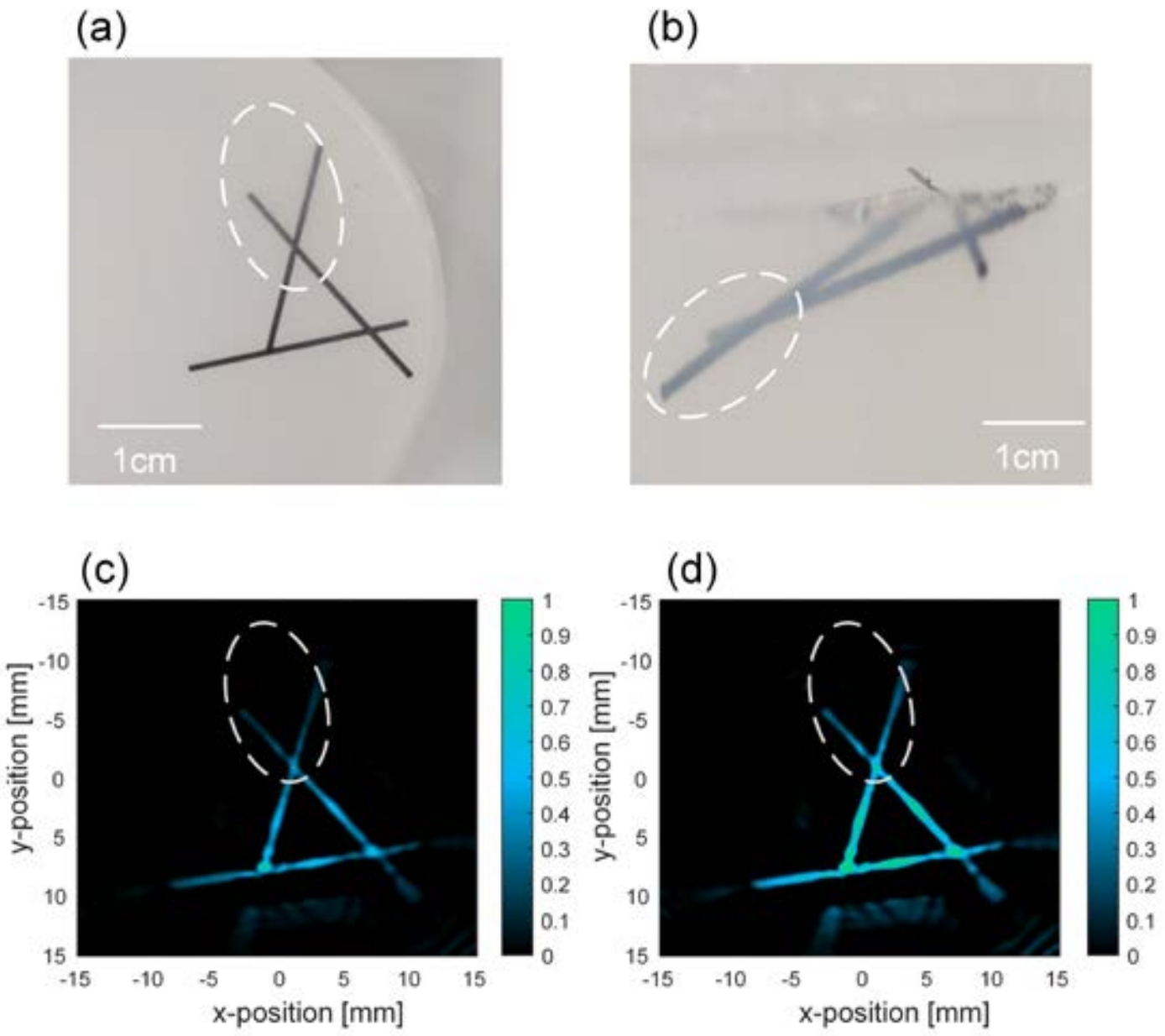

Fig. 9. (a) the top view of the phantom; (b) the lateral view of the phantom; (c) the PAT imaging result with conventional signals sampling method; (d) the PAT imaging result with 4-in-1 PADL module sampling method.

# IV. Conclusion

In this paper, the proposed 4-in-1 PADL module achieves programmable time delay for the PA signals ranging from tens to hundreds of microseconds. Moreover, the FPGA board can conveniently tune the delay time of the PADL module. The delayed signal is restorable in the digital domain with the proposed method according to the delay-line module. By applying the module, one channel DAQ can sample four PA signals concurrently. The PAT system integrated with the 4-in-1 PADL module can reduce the system cost and accelerate imaging speed. By the phantom imaging experiments, the feasibility of the PADL module is well demonstrated. This work provides a potential method to significantly reduce the cost and accelerate the imaging speed of the PAT system[23]. Furthermore, the 4-in-1 PADL module is also appropriate for other kinds of pulse signals' time delay beyond ultrasound. The demonstrated PAT system in this paper, which is based on the 4-in-1 PADL module, has reduced 75% of the DAQ cost or reduced the time consumption of the PAT imaging. By applying more delay line units, the PADL module can merge more pulse signals into one delayed signal that can further reduce the system cost for real-time PA imaging system development.

## V. Acknowledgment

This work was supported by the Natural Science Foundation of Shanghai (18ZR1425000) and the National Natural Science Foundation of China (NSFC) (61805139).